\documentclass[12pt]{article}
\usepackage{epsfig}
\usepackage{amssymb}
\usepackage{amsmath,amsfonts,amssymb,graphicx}
\newcommand {\be}{\begin{equation}}
\newcommand {\ee}{\end {equation}}
\newcommand{\beq}{\begin{eqnarray}}
\newcommand{\eeq}{\end{eqnarray}}

\begin{document}
\title{Review of Neutrino Oscillations With Sterile and Active Neutrinos}
\author{Leonard S. Kisslinger\\
Department of Physics, Carnegie Mellon University, Pittsburgh PA 15213}
\maketitle
\date{}
\noindent
PACS Indices:11.30.Er,14.60.Lm,13.15.+g
\vspace{3mm}


\begin{abstract}
  Recently neutrino oscillation experiments have shown that it is very 
likely that there are one or two sterile neutrinos. In this review neutrino
oscillations with one, two, three sterile and three active neutrinos, and
parameters that are consistent with experiments, are reviewed.
\end{abstract}

\section{Introduction}

This is a review of the method introduced by Sato and collaborators for
three active neutrinos\cite{as97,ks97} extended to three active neutrinos
plus one, two, or three sterile neutrinos. The transition probability for
a neutrino of flavor $f1$ to oscillate to a neutrino of flavor $f2$,
$\mathcal{P}(\nu_{f1} \rightarrow$ $\nu_{f2})$, is derived using S-Matrix theory, 
which is discussed in the next section with $f1,f2 \rightarrow$ muon,
electron neutrinos. 

  In the following three sections the derivation of 
$\mathcal{P}(\nu_\mu \rightarrow$ $\nu_e)$ is described for one, 
two, three sterile neutrinos, with predictions using parameters of four
recent neutrino oscillation experiments. In all three sections a U-matrix
approach is used, introduced with a 3x3 U-martix\cite{as97}, and extended
to a 4x4 U-martix with three active and one sterile neutrino a
5x5 U-martix with three active and two sterile neutrinos and a
6x6 U-martix with three active and three sterile neutrinos.

  From these sections the dependence of the
$\mathcal{P}(\nu_\mu \rightarrow$ $\nu_e)$ neutrino oscillation probability
on the number of sterile neutrinos and oscillation parameters will be shown.

\newpage

\section{$\mathcal{P}(\nu_\mu \rightarrow$ $\nu_e$) for Active
Neutrinos  Derived Using Improved S-Matrix Theory}

Neutrinos are produced as $\nu_f$, with $f$ =flavor=$e, \mu, \tau$. They
do not have definite mass, which is the cause of neutrino oscillatios,
which we now discuss.
Active neutrinos with flavors $\nu_e,\nu_\mu,\nu_ \tau $ are related to
neutrinos with definite mass $\nu_m$, m=1,2,3 by the 3$\times$3 unitary 
matrix, $U$,

\beq
\label{f-mrelation}
      \nu_f &=& U\nu_m \; ,
\eeq
where $\nu_f, \nu_m$ are 3$\times$1 column vectors and $U$ a 3x3 matrix.
Therefore the electron state produced at time t=0 is
\beq
\label{nueto}
       |\nu_e>&=& \sum_{i=1}^{3} U_{1i}|\nu_{m_j}> \; .
\eeq

Making use of quantum theory, a state with energy E satisfies
\beq
\label{Eop}
        i d/dt |E(t)> &=& E |E(t)> {\rm \;\;\;giving} \nonumber \\
          |E(t)> &=& e^{-iEt}|E(t=0)> \; ,
\eeq
or the electron neutrino state of Eq(\ref{nueto}) at time t for the neutrino
at rest (e=m, with $c\equiv 1$) is
\beq
\label{nuett}
  |\nu_e,t>=&=& \sum_{i=1}^{3} U_{1i} e^{-im_i t}|m_i,t=0> {\rm \;\;\; or}
\nonumber \\
     |\nu_e,t>&=& \sum_{i=1}^{3} c_f  |\nu_f> \; ,
\eeq
which shows that an electron neutrino produced at time t=0 oscillates to
neutrinos of different flavors at time t. Therefore an electron neutrino 
produced at t=0 when it travels a distance $L \simeq t$ (as the velocity of 
the very low mass neutrinos is almost the speed of light) oscillates into 
$e, \mu, \tau$ neutrinos. By placing detectors at a distance $L$ this 
oscillation has been measured. The $\nu_\mu$ neutrino has a similar 
relationship, and can also oscillate to a sterile neutrino as we discuss
in sections below.

The 3x3 active neutrino U-matrix is ($sin\theta_{ij} \equiv s_{ij}$, etc).
\vspace{3mm}

$U$=
$\left( \begin{array}{clcr} c_{12}c_{13} & s_{12}c_{13} &s_{13} e^{-i\delta_{CP}}\\
-s_{12}c_{23}-c_{12}s_{23}s_{13}e^{i\delta_{CP}} &c_{12}c_{23}-
s_{12}s_{23}s_{13}e^{i\delta_{CP}}
& s_{23}c_{13} \\ s_{12}s_{23}-c_{12}s_{23}s_{13}e^{i\delta_{CP}} & -c_{12}s_{23}
-s_{12}c_{23}s_{13}e^{i\delta_{CP}} &  c_{23}c_{13} \end{array} \right)$
\vspace{3mm}

\noindent
where $c_{12}=.83,\;s_{12}=.56,\;s_{23}=c_{23}=.7071$, $s_{13}$= 0.19, and
$\delta_{CP}$=0 are used. 

Given the Hamiltonian, H(t), for neutrinos, the neutrino state at time = $t$
is obtained from the state at time = $t_0$ from the S-matrix, $S(t,t_0)$, by
\beq
             |\nu(t)> &=& S(t,t_0)|\nu(t_0)> \\
             i\frac{d}{dt}S(t,t_0) &=& H(t) S(t,t_0) \; .
\eeq
\newpage

In the vacuum the S-matrix is obtained from
\beq
        S_{ab}(t,t_0)&=& \sum_{j=1}^{3} U_{aj} exp^{i E_j (t-t_0)} U^*_{bj} \; ,
\eeq
while for neutrinos travelling through the earth the potential 
$V=1.13 \times 10^{-13}$ ev is included.

The transition probability $\mathcal{P}(\nu_\mu \rightarrow$ $\nu_e$) is 
obtained from the S-Matrix element $S_{12}$:
\beq
\label{PmueS12}
 \mathcal{P}(\nu_\mu \rightarrow \nu_e) &=& (Re[S_{12}])^2 + (Im[S_{12}])^2
\; .
\eeq
\vspace{3mm}

From Ref\cite{khj12}
\beq
\label{ReImS12}
    Re[S12]&=&s_{23}a[cos(\bar{\Delta}L)Im[I_{\alpha^*}]-sin(\bar{\Delta}L)
Re[I_{\alpha^*}]] \nonumber \\
    Im[S12]&=&-c_{23}sin2\theta sin\omega L-s_{23}a[cos(\bar{\Delta}L)
Re[I_{\alpha^*}] \nonumber \\
   &&+sin(\bar{\Delta}L)Im[I_{\alpha^*}] \; ,
\eeq
with $\bar{\Delta}= \Delta-(V+\delta)/2$,  $\Delta=\delta m_{13}^2/(2E)$,
$\delta=\delta m_{12}^2/(2E)$, where the neutrino mass differences are 
$\delta m_{12}^2=7.6x10^{-5}(eV)^2$, $\delta m_{13}^2=2.4x10^{-3}(eV)^2$, 
$sin2\theta= s_{12}c_{12} \frac{\delta}{\omega}$, $a= s_{13}(\Delta-s_{12}^2 
\delta)$, and $E$ is the neutrino energy. Note that $t\rightarrow L$, 
where $L$ is the baseline, for $v_\nu \simeq c$. The neutrino-matter potential
$V=1.13\times 10^{-13}$ eV. 

An important quantity is $I_{\alpha^*}$
\beq
\label{Ialpha*}
    I_{\alpha^*}&=& \int_{0}^{t} dt' \alpha^*(t')e^{-i\bar{\Delta}t'} \; ,
\eeq
with $\alpha(t)= cos(\omega t)-icos2\theta sin(\omega t)$, $\omega=
 \sqrt{\delta^2 +V^2-2\delta V cos(2\theta_{12})}/2$.
In Ref.\cite{khj12}, as in Ref.\cite{ahlo01}, one used 
$\delta,\omega \ll \Delta$ to obtain
\beq
\label{ReImI}
     Re[I_{\alpha^*}]&\simeq& sin\bar{\Delta}L/\bar{\Delta} \nonumber \\
    Im[I_{\alpha^*}]&\simeq& (1-cos\bar{\Delta}L)/\bar{\Delta} \; .
\eeq

In an improved theory\cite{lsk15} it was shown that
\beq
\label{ReImIexact}
     Re[I_{\alpha^*}]&=&  [(\omega-\bar{\Delta}cos2\theta)cos\bar{\Delta} L
sin\omega L \nonumber \\
   &&-(\bar{\Delta}-\omega cos2\theta)sin\bar{\Delta} Lcos\omega L]/
(\omega^2-\bar{\Delta}^2) \\
Im[I_{\alpha^*}]&=& [\bar{\Delta}+\omega cos2\theta-
(\bar{\Delta}+\omega cos2\theta)cos\bar{\Delta} Lcos\omega L \nonumber \\
  &&-(\omega+\bar{\Delta}cos2\theta)sin\bar{\Delta} Lsin\omega L]/(\omega^2
-\bar{\Delta}^2) \nonumber  \; .
\eeq

From Eqs(\ref{ReImS12},\ref{ReImIexact}) $\mathcal{P}(\nu_\mu \rightarrow \nu_e)
= (Re[S_{12}])^2 + (Im[S_{12}])^2$ is obtained, giving the results shown in 
Figure 1.
\clearpage

\hspace{3cm}$\mathcal{P}(\nu_\mu \rightarrow$ $\nu_{e}$) with old and precise 
$I_{\alpha^*}(E,L)$.
\vspace{6cm}

\begin{figure}[ht]
\begin{center}
\epsfig{file=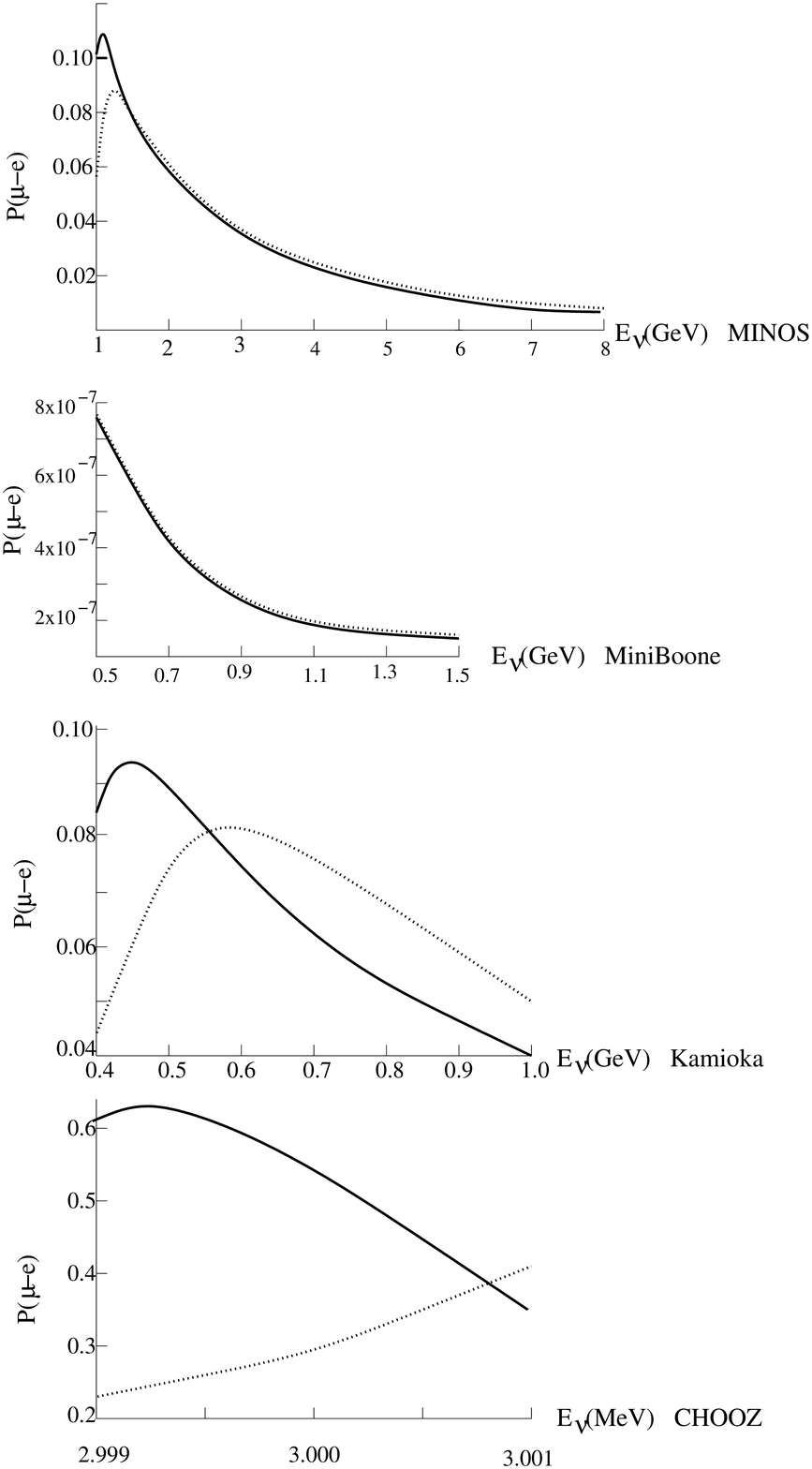,height=12cm,width=10cm}
\end{center}
\caption{$\mathcal{P}(\nu_\mu \rightarrow\nu_e)$ 
for MINOS(L=735 km), MiniBooNE(L=500m), JHF-Kamioka(L=295 km), and 
CHOOZ(L=1.03 km) using the improved 3$\times$3 mixing matrix. Solid curve for 
precise  $I_{\alpha^*}(E,L)$ and dashed curve for approximate  $I_{\alpha^*}(E,L)$,
$s_{13}$=0.19} 
\end{figure}

\newpage

\section{$\mathcal{P}(\nu_\mu \rightarrow$ $\nu_e$) With Three 
Active and One Sterile Neutrino} 

Active neutrinos have only weak and gravitational interactions, and are 
therefore difficult to detect in neutrino oscillation experiments.
Sterile neutrinos have no
interaction except gravity and therefore cannot be detected via the
apparatus used in neutrino oscillation or other experiments. For an
overview of sterile neutrinos and neutrino oscillations see Ref\cite{kmms13}
in which sterile neutrino states are investigated using neutrino oscillation
data. These authors, J. Koppe et. al., considered one and two sterile neutrinos
and discussed both experimental and theoretical publications.

  In the present section we discuss neutrino oscillations with one sterile
neutrino, while in the next section two sterile neutrinos are discussed.

 Motivated  by an experiment measuring neutrino oscillations\cite{mini13}, 
which suggested the existence of at least one sterile neutrino and estimated
the mass differences and mixing angles with active neutrinos, estimates of
$ \mathcal{P}(\nu_\mu \rightarrow\nu_e)$ were made\cite{lsk14}. We now review
this article.

    This is an exension of the method introduced by Sato and 
collaborators for three active neutrino oscillations\cite{as97,ks97}
to three active neutrinos plus one sterile neutrino. Active neutrinos with
flavors $\nu_e,\nu_\mu,\nu_\tau$ and a sterile neutrino $\nu_s$ are 
related to neutrinos with definite mass by
\beq
\label{fm}
      \nu_f &=& U\nu_m \; ,
\eeq
where $U$ is a 4x4 matrix and $\nu_f,\nu_m$ are 4x1 column vectors.
\beq
\label{Uform1}
     U &=& O^{23}\phi O^{13} O^{12} O^{14} O^{24} O^{34} {\rm \;\;with}
\eeq

\hspace{3mm}$O^{23}$=
$\left( \begin{array}{ccclcr} 1 & 0 & 0 & 0 \\ 0 & c_{23} & s_{23} & 0 \\
0 & -s_{23} & c_{23} & 0 \\ 0 & 0 & 0 & 1  \end{array} \right)$, 
\hspace{3mm}
$O^{13}$=
 $\left( \begin{array}{ccclcr} c_{13} & 0 & s_{13} & 0 \\ 0 & 1 & 0 & 0 \\
-s_{13} & 0  & c_{13} & 0 \\ 0 & 0 & 0 & 1  \end{array} \right)$,
\vspace{5mm}

\hspace{3mm}$O^{12}$=
 $\left( \begin{array}{ccclcr} c_{12} & s_{12} & 0 & 0 \\ -s_{12} & c_{12} & 0 & 0 
\\ 0 & 0  & 1 & 0 \\ 0 & 0 & 0 & 1  \end{array} \right)$,
\hspace{3mm}
$O^{14}$=
$\left( \begin{array}{ccclcr} c_\alpha & 0 & 0 & s_\alpha \\ 0 & 1 & 0 & 0 \\
 0 & 0  & 1 & 0 \\ -s_\alpha & 0 & 0 & c_\alpha  \end{array} \right)$,
\vspace{5mm}

\hspace{3mm}$O^{24}$=
$\left( \begin{array}{ccclcr} 1 & 0 & 0 & 0 \\ 0 & c_\alpha & 0 & s_\alpha \\
 0 & 0  & 1 & 0 \\ 0 & -s_\alpha & 0 & c_\alpha  \end{array} \right)$,
\hspace{3mm}
$O^{34}$=
$\left( \begin{array}{ccclcr} 1 & 0 & 0 & 0 \\ 0 & 1 & 0 & 0 \\
 0 & 0  & c_\alpha & s_\alpha \\ 0 & 0 & -s_\alpha & c_\alpha  \end{array} \right)$
\vspace{5mm}

\hspace{3mm}$\phi$=
$\left( \begin{array}{ccclcr} 1 & 0 & 0 & 0 \\ 0 & 1 & 0 & 0 \\
 0 & 0  & e^{i\delta_{CP}} & 0 \\ 0 & 0 & 0 & 1  \end{array} \right)$

\newpage
\noindent
with $c_{12}=.83,\;s_{12}=.56,\;s_{23}=c_{23}=.7071$. We use 
$s_{13}=.15$ from the Daya Bay Collaboration\cite{DB3-7-12}.
In our present work we assume the angles $\theta_{j 4} \equiv \alpha $ for
all three j, and $s_\alpha,c_\alpha=sin\alpha,cos\alpha$. An important aspect of 
this work was to find the dependence of neutrino oscillation probabilities
on $s_\alpha,c_\alpha$.

From Eq(\ref{Uform1}) the 4x4 $U$ matrix is
\vspace{2mm}
\small

$\left( \begin{array}{ccclcr} c_{12}c_{13}c_\alpha &c_{13}(s_{12}c_\alpha-c_{12}s_\alpha^2)& 
- c_{13}s_\alpha^2(c_{12}c_\alpha+s_{12})+s_{13}c_\alpha&  c_{13}s_\alpha c_\alpha
(c_{12}c_\alpha+s_{12})+s_{13}s_\alpha \\ A c_\alpha& -As_\alpha^2+Bc_\alpha &
 -As_\alpha^2c_\alpha-Bs_\alpha^2+c_{13}s_{23}e^{i\delta_{CP}}c_\alpha &
As_\alpha c_\alpha^2+Bs_\alpha c_\alpha+c_{13}s_{23}e^{i\delta_{CP}}s_\alpha\\
Cc_\alpha & -Cs_\alpha^2+Dc_\alpha & -Cs_\alpha^2c_\alpha-Ds_\alpha^2+
c_{13}c_{23}e^{i\delta_{CP}}c_\alpha &
Cs_\alpha c_\alpha^2+Ds_\alpha c_\alpha+c_{13}s_{23}e^{i\delta_{CP}}s_\alpha\\
-s_\alpha & -s_\alpha c_\alpha & -s_\alpha c_\alpha^2 &  c_\alpha^3  \end{array} 
\right)$
\vspace{2mm}
\normalsize
with
\vspace{5mm}
\beq
\label{Uparameters}
   A&=& -(c_{23}s_{12}+c_{12}s_{13}s_{23}e^{i\delta_{CP}}) \nonumber \\
   B&=& (c_{23}c_{12}-s_{12}s_{13}s_{23}e^{i\delta_{CP}}) \\
   C&=& (s_{23}s_{12}-c_{12}s_{13}c_{23}e^{i\delta_{CP}}) \nonumber \\
   D&=& -(s_{23}c_{12}+s_{12}s_{13}c_{23}e^{i\delta_{CP}}) \nonumber \; .
\eeq

Using the formalism of Refs.\cite{as97,ks97} extended to four neutrinos,
the transition probability $ \mathcal{P}(\nu_\mu \rightarrow\nu_e)$ is
obtained from the 4x4 U matrix and the neutrino mass differences
$\delta m_{ij}^2=m_i^2-m_j^2$ for a neutrino beam with energy $E$ and baseline
$L$ by\cite{as97}
\beq
\label{Pue-1}
 \mathcal{P}(\nu_\mu \rightarrow\nu_e) &=& \sum_{i=1}^{4}\sum_{j=1}^{4}
U_{1i}U^*_{1j}U^*_{2i}U_{2j} e^{-i(\delta m_{ij}^2/E)L} \; ,
\eeq
or, since with $\delta_{CP}=0$ $U^*_{ij}=U_{ij}$,
\beq
\label{Pue2}
\mathcal{P}(\nu_\mu \rightarrow\nu_e) &=& U_{11}U_{21}[ U_{11}U_{21}+
 U_{12}U_{22} e^{-i\delta L}+ U_{13}U_{23} e^{-i\Delta L}+
U_{14}U_{24} e^{-i\gamma L}]+ \nonumber \\
  &&  U_{12}U_{22}[ U_{11}U_{21}e^{-i\delta L}+ U_{12}U_{22} + U_{13}U_{23} 
e^{-i\Delta L}+U_{14}U_{24} e^{-i\gamma L}]+ \nonumber \\
  &&  U_{13}U_{23}[ U_{11}U_{21}e^{-i\Delta L}+ U_{12}U_{22}e^{-i\Delta L}
 + U_{13}U_{23} +U_{14}U_{24} e^{-i\gamma L}]+ \nonumber \\
   &&  U_{14}U_{24}[(U_{11}U_{21}+ U_{12}U_{22} + U_{13}U_{23})e^{-i\gamma L}
 +U_{14}U_{24}] \; ,
\eeq 
with $\delta=\delta m_{12}^2/2E,\; \Delta=\delta m_{13}^2/2E,\; \gamma=
\delta m_{j4}^2/2E$ (j=1,2,3). The neutrino mass differences are 
$\delta m_{12}^2=7.6 \times 10^{-5}(eV)^2$, $\delta m_{13}^2 = 2.4\times 10^{-3} 
(eV)^2$; and we use both $\delta m_{j4}^2=0.9 (eV)^2$ and 
$\delta m_{j4}^2=0.043 (eV)^2$, since $\delta m_{j4}^2=0.043 (eV)^2$ was the
best fit parameter found via the 2013 MiniBooNE analysis, while
$\delta m_{j4}^2=0.9 (eV)^2$ is the best fit using the 2013 MiniBooNE data
and previous experimental fits\cite{mini13}.

  Note that in Refs\cite{hjk11,khj12} $\mathcal{P}(\nu_\mu \rightarrow\nu_e)=
|S_{12}|^2$, with $S_{12}$ obtained from the 3x3 U-matrix and the $\delta m_{ij}$
parameters. Therefore our formalism, given by Eq(\ref{Pue2}), is quite 
different, and as will be shown for the same $L,E$ the magnitude of
$\mathcal{P}(\nu_\mu \rightarrow\nu_e)$ is also different. Since the S-matrix
formalism was not used in Refs.\cite{as97,ks97} for the 3x3 study, 
$\mathcal{P}(\nu_\mu \rightarrow\nu_e)$ was quite different from
Refs\cite{hjk11,khj12}. 
\newpage

From Eq(\ref{Pue2}),
\beq
\label{Pue}
 \mathcal{P}(\nu_\mu \rightarrow\nu_e) &=& U_{11}^2 U_{21}^2+
 U_{12}^2 U_{22}^2+ U_{13}^2 U_{23}^2+  \nonumber \\
  && U_{14}^2 U_{24}^2+ 2U_{11} U_{21} U_{12} U_{22} cos\delta L + \\
  && 2(U_{11} U_{21} U_{13} U_{22}+ U_{12} U_{22} U_{13} U_{23})cos\Delta L+
\nonumber \\
  &&2U_{14}U_{24}(U_{11} U_{21}+U_{12} U_{22}+U_{13} U_{23})cos\gamma L 
\nonumber \; .
\eeq 

Using the parameters given above,
\beq
\label{Uij}
   U_{11}&=& .822 c_\alpha {\rm \;\;\;}U_{12}=-.554s_\alpha^2 +0.084c_\alpha 
\nonumber \\
   U_{13}&=&-.822s_\alpha^2c_\alpha-.554s_\alpha^2+.15c_\alpha {\rm \;\;\;}
U_{14}=.822s_\alpha c_\alpha^2+.554s_\alpha c_\alpha+.15s_\alpha \nonumber \\
   U_{21}&=& -.484c_\alpha {\rm \;\;\;} U_{22}=.484s_\alpha^2 +
.527 c_\alpha \\
    U_{23}&=& .484c_\alpha-.527s_\alpha^2+.7c_\alpha {\rm \;\;\;}
U_{24}=-.484 s_\alpha c_\alpha^2+.527s_\alpha c_\alpha+.7s_\alpha \nonumber \; .
\eeq
With the addition of a sterile neutrino, the 4th neutrino, there are three
new angles, $\theta_{14}$, $\theta_{24}$, and $\theta_{34}$. The main
assumption is that these three angles are the same, $\theta_{j4}=\alpha$.
The angle $\alpha$ is the main parameter that is being studyied.

Two values for the sterile-active mass differences are used. The 
most widely accepted value for $m_4^2-m_1^2$ is $0.9 (eV)^2$\cite{mini13},
but we also use $m_4^2-m_1^2=.043 (eV)^2$ from the 2013 MiniBoonE result
to test the sensitivity of $\mathcal{P}(\nu_\mu \rightarrow \nu_e)$ to
the sterile neutrino-active neutrinos mass differences. Since 
$m_4^2-m_1^2>>m_j^2-m_i^2$ for (i,j)=1,2,3, we assume that $m_4^2-m_j^2=
m_4^2-m_1^2$.

Figure 2 shows the results for $\mathcal{P}(\nu_\mu \rightarrow \nu_e)$
for the four experiments with $m_4^2-m_1^2=0.9 (eV)^2$ and $\alpha=45^o,
60^o,30^o$. As one can see, $\mathcal{P}(\nu_\mu \rightarrow \nu_e)$ is very 
strongly dependent on $\alpha$.

Next $m_4^2-m_1^2=0.043 (eV)^2$ was used, as found in the recent MiniBooNE
experiment, to study the effects of $m_4^2-m_1^2$ on 
$\mathcal{P}(\nu_\mu \rightarrow \nu_e)$, with $\alpha=45^o,30^o,60^o$, as 
shown in Figure 3.

Note for $\alpha=0$ (no sterile-active mixing) $U_{14}=0$. Therefore,
$\mathcal{P}(\nu_\mu \rightarrow \nu_e)$ is a 3x3 theory;
however, we find that $\mathcal{P}(\nu_\mu \rightarrow \nu_e)$ is different 
with the model of Refs.\cite{as97,ks97}.

An article Neutrino Oscillations With Recently Measured Sterile-Active Neutrino 
Mixing Angle $sin(\alpha) = 0.16$ was recently published\cite{lskz15}, with 
estimates of $\mathcal{P}(\nu_\mu \rightarrow \nu_e)$, is shown in Figure 4.

\clearpage
\setlength{\topmargin}{0.25in}
\setlength{\textheight}{8.5in}

\begin{figure}[ht]
\begin{center}
\epsfig{file=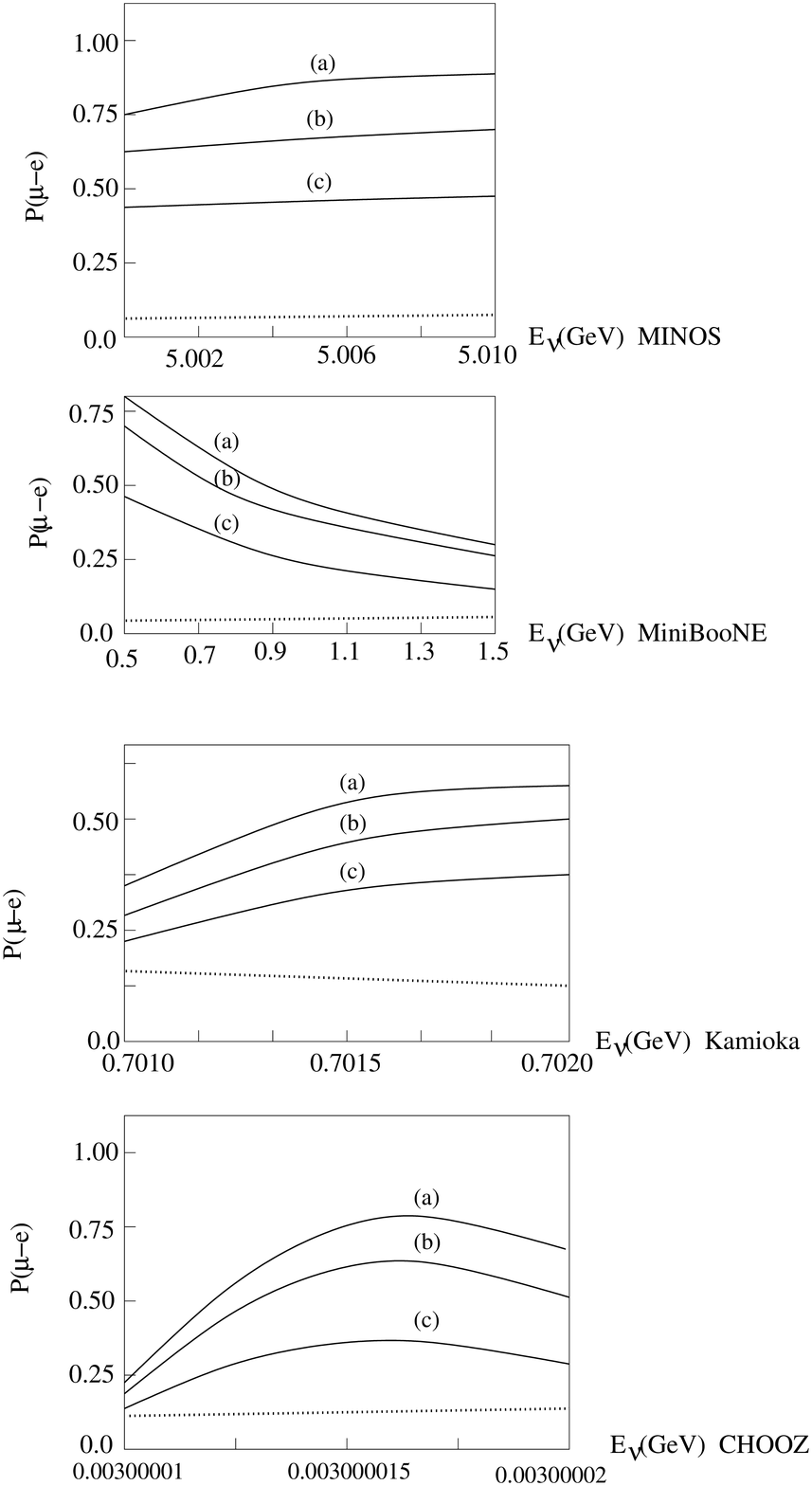,height=18cm,width=12cm}
\end{center}
\caption{\hspace{5mm} The ordinate is $\mathcal{P}(\nu_\mu \rightarrow\nu_e)$ 
for MINOS(L=735 km), MiniBooNE(L=500m), JHF-Kamioka(L=295 km), and 
CHOOZ(L=1.03 km) using the 4x4 U matrix with 
$\delta m_{j4}^2=0.9 (eV)^2$, $s_{13}=0.15$, and (a),(b),(c) for $\alpha=45^o, 
60^o, 30^o$. The dashed curves are for $\alpha=0$ (3x3).}
\end{figure}

\clearpage

\begin{figure}[ht]
\begin{center}
\epsfig{file=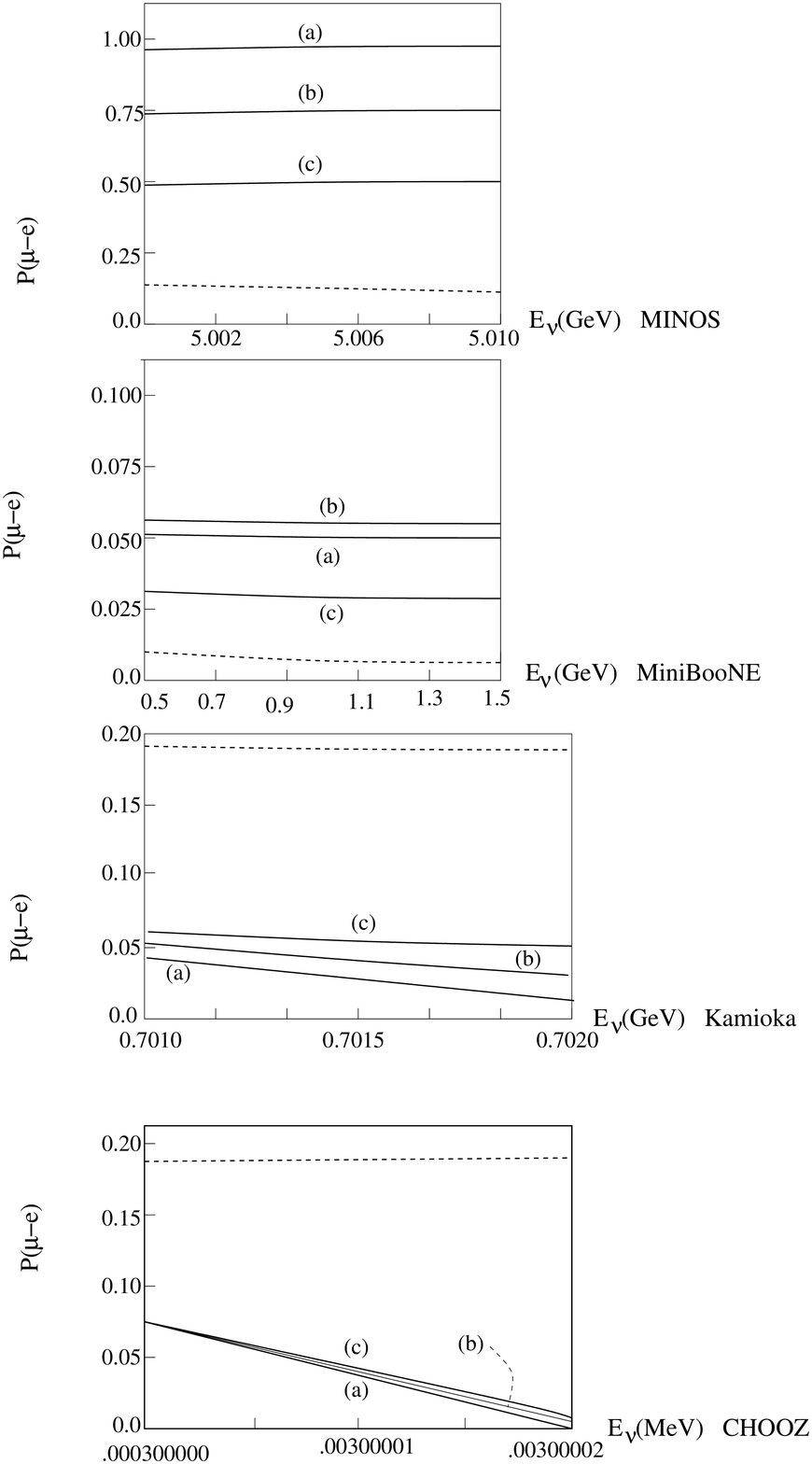,height=18cm,width=12cm}
\end{center}
\caption{\hspace{5mm} The ordinate is $\mathcal{P}(\nu_\mu \rightarrow\nu_e)$ 
for MINOS(L=735 km), MiniBooNE(L=500m), JHF-Kamioka(L=295 km), and 
CHOOZ(L=1.03 km) using the 4x4 U matrix with $\delta m_{j4}^2=0.043 (eV)^2$,
 $s_{13}=0.15$, and (a),(b),(c) for $\alpha=45^o, 60^o, 30^o$. The dashed curves
are for $\alpha=0$ (3x3).}
\end{figure}
\clearpage

\begin{figure}[ht]
\begin{center}
\epsfig{file=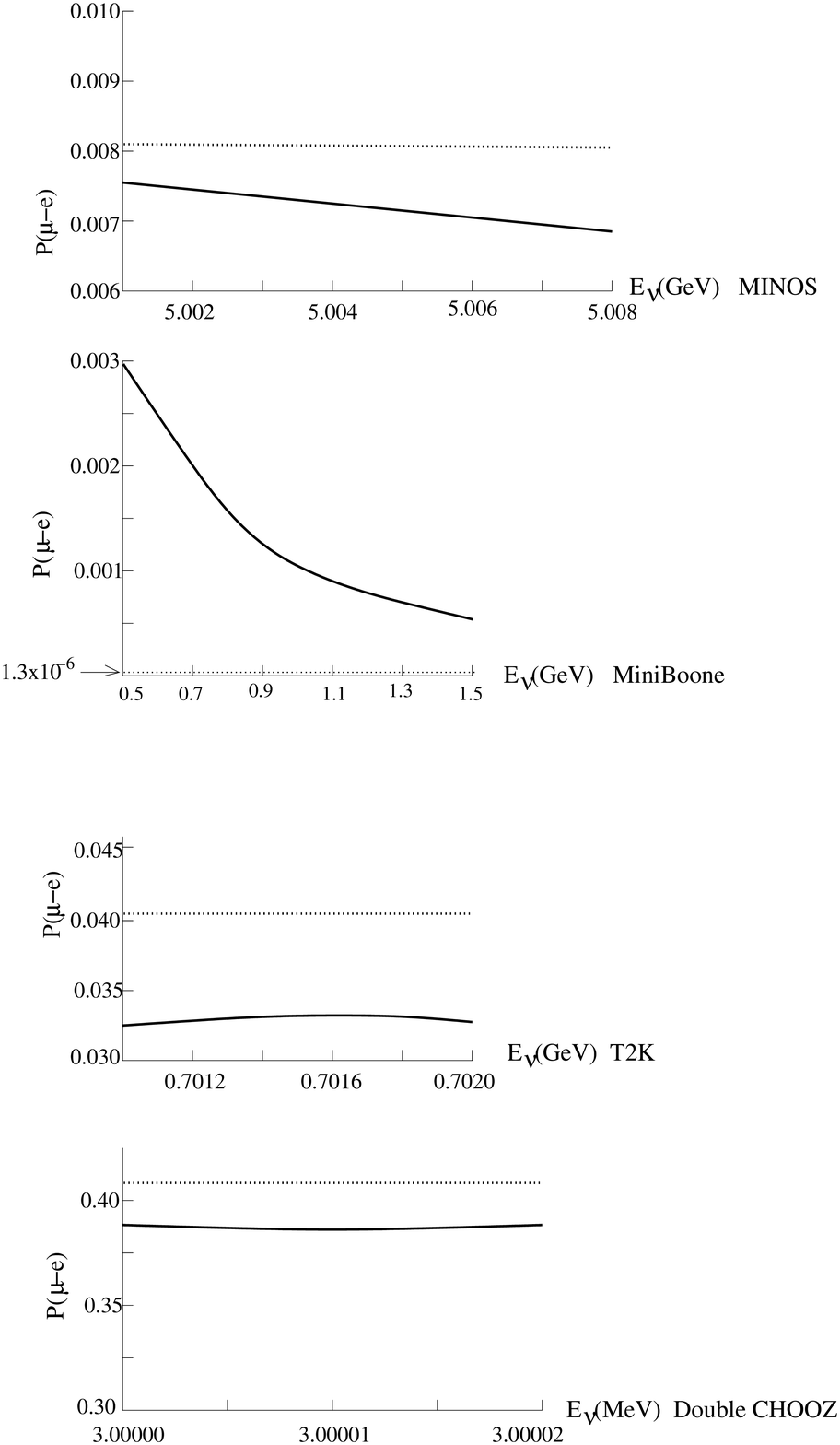,height=18cm,width=12cm}
\end{center}
\caption{\hspace{5mm} The ordinate is $\mathcal{P}(\nu_\mu \rightarrow\nu_e)$ 
for MINOS(L=735 km), MiniBooNE(L=500m), JHF-Kamioka(L=295 km), and 
CHOOZ(L=1.03 km) using the 4x4 U matrix with $\delta m_{j4}^2=0.9 (eV)^2$
and $sin(\alpha)\simeq 0.16$. The dashed curves are for $\alpha=0$ (3x3).}
\end{figure}
\clearpage 
\section{$\mathcal{P}(\nu_\mu \rightarrow$ $\nu_e$) With Three 
Active and Two Sterile Neutrinos} 
 
Recent reviews of experimental data on neutrino oscillations\cite{kmms13,
ggllz15} find that there probably are two sterile neutrino with the mass and 
mixing angles used in the present review, which is based on Ref\cite{lskzz15}.

  With three active and two sterile neutrinos one uses a 5x5 U-matrix to
estimate the transition porbability for a muon neutrino to oscillate to
an electron neutrino $\mathcal{P}(\nu_\mu \rightarrow$ $\nu_e)$. The U-matrix
that relates neutrinos with definite mass m to neutrinos with definite flavor
f, Eq({\ref{fm}), with the sterile-active mixing angles $s_\alpha=sin(\alpha), 
c_\alpha=cos(\alpha), s_\beta=sin(\beta),c_\beta= cos(\beta)$, where 
$\alpha=\theta_{i4},\beta=\theta_{i5}$ are sterile-active 
neutrino mixing angles, with i=1,2,3, and $\delta_{CP}$=0, is

\beq
\label{Uform2}
    U &=& O^{23}O^{13} O^{12} O^{14} O^{24} O^{34} O^{15} O^{25} O^{35} 
O^{45} \; , 
\eeq
where ($O^{45}$, giving sterile-sterile neutrino mixing, is not shown)
\vspace{3mm}

$O^{23}$=
 $\left( \begin{array}{ccclcr} 1 & 0 & 0 & 0 & 0 \\ 0 & c_{23} & s_{23} & 0 
& 0\\
0 & -s_{23} & c_{23} & 0 & 0\\ 0 & 0 & 0 & 1 & 0\\ 0 & 0 & 0 & 0 & 1 \end{array} 
\right)$
\hspace{3mm}$O^{13}$=
$\left( \begin{array}{ccclcr} c_{13} & 0 & s_{13} & 0 & 0\\ 0 & 1 & 0 & 0 
& 0 \\-s_{13} & 0  & c_{13} & 0 & 0 \\ 0 & 0 & 0 & 1 & 0\\ 0 & 0 & 
0 & 0 & 1  \end{array} \right)$
\vspace{3mm}

$O^{12}$=
 $\left( \begin{array}{ccclcr} c_{12} & s_{12} & 0 & 0 & 0\\ -s_{12} & 
c_{12} & 0 & 0 & 0 \\ 0 & 0  & 1 & 0 & 0 \\ 0 & 0 & 0 & 1 & 0\\
0 & 0 & 0 & 0 & 1 \end{array} \right)$
\hspace{3mm}$O^{14}$=
 $\left( \begin{array}{ccclcr} c_\alpha & 0 & 0 & s_\alpha & 0\\ 
 0 & 1  & 0 & 0 & 0\\  0 & 0 & 1 & 0 & 0\\
 -s_\alpha & 0 & 0 & c_\alpha & 0\\
0& 0 & 0 & 0 & 1 \end{array} \right)$
\vspace{3mm}

$O^{24}$=
$ \left( \begin{array}{ccclcr} 1 & 0 & 0 & 0 & 0 \\ 0 & c_\alpha & 0 & 
s_\alpha & 0\\ 0 & 0 & 1 & 0 & 0\\ 0 & -s_\alpha & 0 & c_\alpha & 0 \\
 0 & 0  & 0 & 1 & 0\\
0& 0 & 0 & 0 & 1 \end{array} \right)$
\hspace{3mm}$O^{34}$=
$\left( \begin{array}{ccclcr} 1 & 0 & 0 & 0 & 0 \\ 0 & 1 & 0 & 0 & 0 \\
 0 & 0  & c_\alpha & s_\alpha & 0 \\ 0 & 0 & -s_\alpha & c_\alpha & 0\\
0& 0 & 0 & 0 & 1 \end{array} \right)$

\vspace{3mm}
$\phi$=
$\left( \begin{array}{ccclcr} c_\beta & 0 & 0 & 0 & s_\beta\\
 0 & 1 & 0 & 0 & 0\\ 0 & 0 & 1 & 0 & 0\\ 0 & 0 & 0 & 1 & 0\\ 
- s_\beta & 0 & 0 & 0 & c_\beta \end{array} 
\right)$
\hspace{3mm}$O^{25}$=
 $\left( \begin{array}{ccclcr}  1 & 0 & 0 & 0 & 0\\ 0 & c_\beta & 0 & 0 &
 s_\beta\\ 0 & 0 & 1 & 0 & 0\\ 0 & 0 & 0 & 1 & 0 \\ 
 0 &-s_\beta  & 0 & 0 & c_\beta \end{array} \right)$
\vspace{3mm}

\hspace{3cm}$O^{35}$=
$\left( \begin{array}{ccclcr} 1 & 0 & 0 & 0 & 0\\ 0 & 1 & 0 & 0 & 0\\
 0 & 0 & c_\beta & 0 & s_\beta\\ 0 & 0 & 0 & 1 & 0\\ 0& 0 
 & -s_\beta  & 0 & c_\beta\end{array} \right)$

\vspace{5mm}

  $ \mathcal{P}(\nu_\mu \rightarrow\nu_e)$ 
is obtained from the 5x5 U matrix and the neutrino mass differences
$\delta m_{ij}^2=m_i^2-m_j^2$ for a neutrino beam with energy $E$ and baseline
$L$ by
\beq
\label{Puez-1}
 \mathcal{P}(\nu_\mu \rightarrow\nu_e) &=& Re[\sum_{i=1}^{5}\sum_{j=1}^{5}
U_{1i}U^*_{1j}U^*_{2i}U_{2j} e^{-i(\delta m_{ij}^2/E)L}] \; .
\eeq

From Eq(\ref{Uform2}), multiplying the nine 5x5 $O$ matrices, one obtains the 
matrix U. With $\delta_{CP}$=0, $U^*_{ij}=U_{ij}$, so we only need $U_{1j},U_{2j}$.
The active neutr1no mixing parameters\cite{hjk11} are $c23=s23=0.7071,
c13=0.989,s13=0.15,c12=0.83,s12=0.56$.
\beq
\label{U1j}
  U_{11}&=&.821 ca{\rm \;}cb  \nonumber \\
  U_{12} &=& (.554 ca - .821 sa^2) cb - .821 ca{\rm \;}sb^2  \nonumber \\
 U_{13}&=&(.15 ca-.554 sa^2-.821ca{\rm \;}sa^2)cb-(.554 ca - .821 sa^2)sb^2
\nonumber \\
      && +.821 ca{\rm \;}cb{\rm \;}sb^2\nonumber \\
 U_{14} &=&cb(.15sa +.554 ca{\rm \;}sa + .821 ca^2{\rm \;}sa)-.821ca{\rm \;}
cb^2{\rm \;}sb^2
\nonumber \\
   && -(.554 ca-.821 sa^2)cb{\rm \;}sb^2-(.15 ca-.554 sa^2-.821 ca sa^2)sb^2
\nonumber \\
U_{15} &=&.821ca{\rm \;}sb{\rm \;}cb^3+(.15sa+.554ca{\rm \;}sa+
.821 ca^2{\rm \;}sa)sb \nonumber \\
   && +(.554 ca-.821 sa^2)cb^2{\rm \;}sb+(.15ca-.554sa^2-.821ca{\rm \;}sa^2)
cb{\rm \;}sb \nonumber \\ 
 U_{21}&=&-.484ca{\rm \;}cb \\
  U_{22}&=&(.527ca+.484 sa^2)cb+.484ca{\rm \;}sb^2) \nonumber \\
  U_{23}&=& (.699ca-.527sa^2+.484ca{\rm \;}sa^2)cb-(.527ca+.484sa^2)sb^2 
+.484ca{\rm \;}cb{\rm \;}sb^2\nonumber \\
  U_{24}&=& cb(.699 sa+.527ca{\rm \;}sa-.484ca^2{\rm \;}sa)+
.484ca{\rm \;}cb^2{\rm \;}sb^2 \nonumber \\
   && -(.527ca +.484sa^2)cb{\rm \;}sb^2 -(.699ca-.527sa^2+.484ca{\rm \;}sa^2) 
sb^2 \nonumber \\
  U_{25}&=& -.484 ca{\rm \;}sb{\rm \;}cb^3 +(.699sa+.527ca{\rm \;}sa
-.484ca^2{\rm \;}sa)sb\nonumber \\
   && +(.527ca+.484sa^2)cb^2{\rm \;}sb +(.699ca -.527sa^2+.484 ca{\rm \;}sa^2) 
cb{\rm \;}sb
\nonumber
\eeq

  The active neutrino mass differences are $\delta m_{12}^2=m_2^2-m_1^2=
7.6 \times 10^{-5}(eV)^2$, $\delta m_{13}^2=m_3^2-m_1^2\simeq \delta m_{23}^2 = 
2.4\times 10^{-3} (eV)^2$. From Ref\cite{kmms13} the two sterile-active 
mass differences are  $\delta m_{4i}^2=m_4^2-m_i^2 \simeq$ 0.5 (eV)$^2$ ,
$\delta m_{5i}^2= m_5^2-m_i^2 \simeq$ 0.9 (eV)$^2$, with i=1,2,3 for active 
neutrinos, and $\delta m_{54}^2= m_5^2-m_4^2 \simeq$ 0.4 (eV)$^2$
\newpage

  For $\mu$ neutrino disappearance the sterile-active neutrino mixing angle
$\theta_{\mu\mu}$ is given by\cite{kmms13} $sin2\theta_{\mu\mu}=2|U_{\mu4}|
\sqrt{1-U_{\mu4}^2}$, with a similar form for e-neutrino disappearance. From 
Ref\cite{kmms13} $U_{\mu4}\simeq U_{e4}\simeq U_{\mu5}\simeq U_{e5}=.13-.17$.
Therefore $\alpha \simeq \beta \simeq 7.5^o$ to $10^o$, with $9.2^o$ the 
sterile-active mixing angle used in Refs\cite{lsk14,lsk15}.

With the mass differences $\delta m_{12}^2$, $\delta m_{13}^2$, $\delta m_{23}^2$,
 $\delta m_{4i}^2$, $\delta m_{5i}^2$, $\delta m_{54}^2$ given above, and the
definitions $\delta=\delta m_{12}^2/2E$ , $\Delta=\delta m_{13}^2/2E$, $\gamma= 
\delta m_{4i}^2/2E$, $\lambda= \delta m_{5i}^2/2E$, $\kappa=\delta m_{54}^2/2E$,

\beq
\label{Pue3}
\mathcal{P}(\nu_\mu \rightarrow \nu_{e}) &=&Re[U_{11}U_{21}( U_{11}U_{21}+
 U_{12}U_{22} e^{-i\delta L}+ U_{13}U_{23} e^{-i\Delta L}+ \nonumber \\
  && U_{14}U_{24} e^{-i\gamma L}+U_{15}U_{25} e^{-i\lambda L})+  \\
  &&  U_{12}U_{22}( U_{11}U_{21}e^{-i\delta L}+ U_{12}U_{22} + U_{13}U_{23} 
e^{-i\Delta L}+ \nonumber\\
  && U_{14}U_{24}e^{-i\gamma L}+U_{15}U_{25} e^{-i\lambda L})+ 
  U_{13}U_{23}( U_{11}U_{21}e^{-i\Delta L}+ U_{12}U_{22}e^{-i\Delta L}
 \nonumber \\
  && + U_{13}U_{23}+ U_{14}U_{24}e^{-i\gamma L}+U_{15}U_{25} e^{-i\lambda L}) 
    +U_{14}U_{24}((U_{11}U_{21}+U_{12}U_{22}
\nonumber\\
 && + U_{13}U_{23})e^{-i\gamma L}+U_{14}U_{24}+U_{15}U_{25}e^{-i\kappa L})
\nonumber \\
  &&+U_{15}U_{25}((U_{11}U_{21}+U_{12}U_{22}+ U_{13}U_{23}) e^{-i\lambda L}+
U_{14}U_{24}e^{-i\kappa L}+ U_{15}U_{25})] \nonumber
\eeq 
\vspace{3mm}

 From Eq(\ref{Pue3})
\beq
\label{Pue4}
 \mathcal{P}(\nu_\mu \rightarrow\nu_e) &=& U_{11}^2 U_{21}^2+
 U_{12}^2 U_{22}^2+ U_{13}^2 U_{23}^2+ U_{14}^2 U_{24}^2+U_{15}^2 U_{25}^2 + 
\nonumber \\
  &&  2U_{11} U_{21} U_{12} U_{22} cos\delta L + \\
  && 2(U_{11} U_{21} U_{13} U_{23}+ U_{12} U_{22} U_{13} U_{23})cos\Delta L+
\nonumber \\
  &&2U_{14}U_{24}(U_{11} U_{21}+ U_{12} U_{22}+ U_{13} U_{23})cos\gamma L+
\nonumber \\
  && 2U_{15}U_{25}(U_{11} U_{21}+ U_{12} U_{22}+ U_{13} U_{23})cos\lambda L+
 \nonumber \\
  && 2 U_{14}U_{24}U_{15}U_{25} cos\kappa L \nonumber \; .
\eeq 
\vspace{3mm}

  From Eq(\ref{Pue2}) and the discussion below that equation, $\alpha
\simeq \beta \simeq 7.5^o$, with $sa=sb \simeq 0.131$ and $ca=cb\simeq 0.991$,
and $\alpha\simeq \beta \simeq 10^o$, with $sa=sb \simeq 0.174$ and 
$ca=cb\simeq 0.985$, which are used to determine $U_{1j},U_{2j}$ in Eq(\ref{U1j}).

 In Fig. 5 the results of the two
sterile neutrinos on $\mathcal{P}(\nu_\mu \rightarrow \nu_e)$ using 
Eq(\ref{Pue4}) and the parameters obtained from Refs\cite{kmms13,ggllz15}
are shown for four experimental neutrino oscillation experiments. 

The figure also shows $\mathcal{P}(\nu_\mu \rightarrow \nu_e)$ with 
$\alpha=\beta= 0^o$, giving the results of a recent 3x3 S-mtrix 
calculation\cite{lsk15} to compare to the results with two sterile neutrinos.

\clearpage
\hspace{3cm}Using Eq(\ref{Pue4}), one finds $\mathcal{P}(\nu_\mu \rightarrow 
\nu_e)$
\vspace{5.7cm}

\begin{figure}[ht]
\begin{center}
\epsfig{file=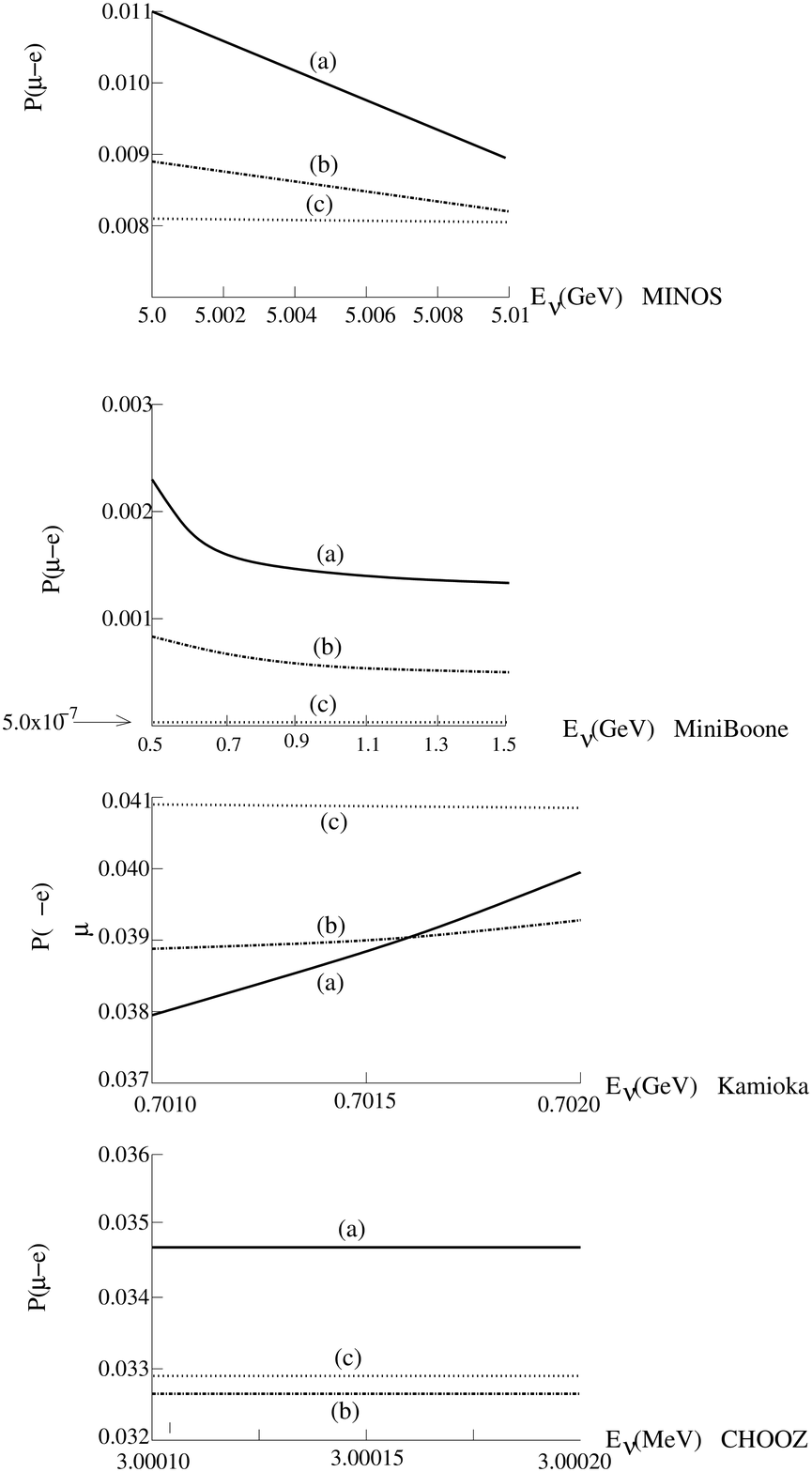,height=12cm,width=10cm}
\end{center}
\caption{$\mathcal{P}(\nu_\mu \rightarrow\nu_e)$ 
for MINOS(L=735 km), MiniBooNE(L=500m), JHF-Kamioka(L=295 km), and 
CHOOZ(L=1.03 km). (a) solid  $\alpha=\beta$=$10.0^o$;  
(b) dash-dotted for $\alpha=\beta$ =$7.5^o$; 
(c) dashed curve for $\alpha=\beta=\gamma$=$0^o$ giving the 3x3 result .}
\end{figure}
\newpage

\section{$\mathcal{P}(\nu_\mu \rightarrow$ $\nu_e$) With Three 
Active and Three Sterile Neutrinos}

This review of neutrino oscillations with three active and three sterile
neutrinos is based on Ref\cite{lskzzz15}. Active neutrinos with flavors 
$\nu_e,\nu_\mu,\nu_\tau$ and three sterile 
neutrinos, $\nu_{s_1},\nu_{s_2},\nu_{s_3}$ are related to neutrinos with 
definite mass by
\beq
\label{f-mrelation}
      \nu_f &=& U\nu_m \; ,
\eeq
where $U$ is a 6x6 matrix and $\nu_f,\nu_m$ are 6x1 column vectors.
We use the notation  $s_{ij}, c_{ij}=sin\theta_{ij},cos\theta_{ij}$, with 
$\theta_{12}, \theta_{23}, \theta_{13}$ the mixing angles for active neutrinos; 
and $s_\alpha=sin(\alpha), c_\alpha=cos(\alpha), s_\beta=sin(\beta)$, etc, where 
$\alpha,\beta,\gamma$ are sterile-active neutrino mixing angles.  

\beq
\label{Uform3}
     U &=& O^{23}O^{13} O^{12} O^{14} O^{24} O^{34} O^{15} O^{25} O^{35} 
O^{45} O^{16}O^{26} O^{36} O^{46} O^{56} 
\eeq
with ($O^{45}$, $O^{46}$, and $O^{56}$, giving sterile-sterile neutrino
mixing, are not shown)
\vspace{3mm}

$O^{23}$=
 $\left( \begin{array}{ccclcr} 1 & 0 & 0 & 0 & 0 & 0 \\ 0 & c_{23} & s_{23} & 0 
& 0 & 0 \\
0 & -s_{23} & c_{23} & 0 & 0 & 0 \\ 0 & 0 & 0 & 1 & 0 & 0 \\ 0 & 0 & 0 & 0 & 1 &
 0 \\ 0 & 0 & 0 & 0 & 0 & 1 \end{array} \right)$
\hspace{3mm}$O^{13}$=
$\left( \begin{array}{ccclcr} c_{13} & 0 & s_{13} & 0 & 0 & 0 \\ 0 & 1 & 0 & 0 
& 0 & 0 \\-s_{13} & 0  & c_{13} & 0 & 0 & 0 \\ 0 & 0 & 0 & 1 & 0 & 0\\ 0 & 0 & 
0 & 0 & 1 & 0\\  0 & 0 & 0 & 0 & 0 & 1  \end{array} \right)$
\vspace{3mm}

$O^{12}$=
 $\left( \begin{array}{ccclcr} c_{12} & s_{12} & 0 & 0 & 0 & 0\\ -s_{12} & 
c_{12} & 0 & 0 & 0 & 0 \\ 0 & 0  & 1 & 0 & 0 & 0 \\ 0 & 0 & 0 & 1 & 0 & 0\\
0 & 0 & 0 & 0 & 1 & 0\\ 0 & 0 & 0 & 0 & 0 & 1 \end{array} \right)$
\hspace{3mm}$O^{14}$=
 $\left( \begin{array}{ccclcr} c_\alpha & 0 & 0 & s_\alpha & 0 & 0\\ 
 0 & 1  & 0 & 0 & 0 & 0\\  0 & 0 & 1 & 0 & 0 & 0\\
 -s_\alpha & 0 & 0 & c_\alpha & 0 & 0\\
0& 0 & 0 & 0 & 1 & 0 \\ 0 & 0 & 0 & 0 & 0 & 1 \end{array} \right)$
\vspace{3mm}

$O^{24}$=
$ \left( \begin{array}{ccclcr} 1 & 0 & 0 & 0 & 0 & 0 \\ 0 & c_\alpha & 0 & 
s_\alpha & 0 & 0 \\ 0 & 0 & 1 & 0 & 0 & 0 \\ 0 & -s_\alpha & 0 & c_\alpha & 0 & 0 
 \\
 0 & 0  & 0 & 1 & 0  & 0\\
0& 0 & 0 & 0 & 1 & 0 \\ 0 & 0 & 0 & 0 & 0 & 1  \end{array} \right)$
\hspace{3mm}$O^{34}$=
$\left( \begin{array}{ccclcr} 1 & 0 & 0 & 0 & 0 & 0 \\ 0 & 1 & 0 & 0 & 0 & 0 \\
 0 & 0  & c_\alpha & s_\alpha & 0 & 0 \\ 0 & 0 & -s_\alpha & c_\alpha & 0 & 0\\
0& 0 & 0 & 0 & 1 & 0 \\ 0 & 0 & 0 & 0 & 0 & 1   \end{array} \right)$
\newpage

$O^{15}$=
$\left( \begin{array}{ccclcr} c_\beta & 0 & 0 & 0 & s_\beta & 0\\
 0 & 1 & 0 & 0 & 0 & 0 \\ 0 & 0 & 1 & 0 & 0 & 0 \\ 0 & 0 & 0 & 1 & 0 & 0 \\ 
- s_\beta & 0 & 0 & 0 & c_\beta & 0 \\ 0 & 0 & 0 & 0 & 0 & 1  \end{array} 
\right)$
\hspace{3mm}$O^{25}$=
 $\left( \begin{array}{ccclcr}  1 & 0 & 0 & 0 & 0 & 0 \\ 0 & c_\beta & 0 & 0 &
 s_\beta & 0\\ 0 & 0 & 1 & 0 & 0 & 0 \\ 0 & 0 & 0 & 1 & 0 & 0 \\ 
 0 &-s_\beta  & 0 & 0 & c_\beta & 0 \\ 0 & 0 & 0 & 0 & 0 & 1  \end{array}
\right)$
\vspace{3mm}

$O^{35}$=
$\left( \begin{array}{ccclcr} 1 & 0 & 0 & 0 & 0 & 0 \\ 0 & 1 & 0 & 0 & 0 & 0 \\
 0 & 0 & c_\beta & 0 & s_\beta & 0 \\ 0 & 0 & 0 & 1 & 0 & 0 \\ 0& 0 
 & -s_\beta  & 0 & c_\beta & 0 \\ 0 & 0 & 0 & 0 & 0 & 1 \end{array} \right)$
\hspace{3mm}$O^{16}$=
 $\left( \begin{array}{ccclcr} c_\gamma & 0 & 0 & 0  & 0 & s_\gamma\\
 0 & 1 & 0 & 0 & 0 & 0 \\ 0 & 0 & 1 & 0 & 0 & 0 \\ 0 & 0 & 0 & 1 & 0 & 0 \\
 0 & 0 & 0 & 0 & 1 & 0 \\ - s_\gamma & 0 & 0 & 0 & 0 & c_\gamma   \end{array} 
\right)$
\vspace{3mm}

$O^{26}$=
 $\left( \begin{array}{ccclcr}  1 & 0 & 0 & 0 & 0 & 0 \\ 0 & c_\gamma & 0 & 0 
 & 0 & s_\gamma\\ 0 & 0 & 1 & 0 & 0 & 0 \\ 0 & 0 & 0 & 1 & 0 & 0 \\ 
 0 & 0 & 0 & 0 & 1 & 0 \\ 
 0 &-s_\gamma  & 0 & 0 & 0 & c_\gamma  \end{array}
\right)$
\hspace{3mm}$O^{36}$=
 $\left( \begin{array}{ccclcr} 1 & 0 & 0 & 0 & 0 & 0 \\ 0 & 1 & 0 & 0 & 0 & 0 \\
 0 & 0 & c_\gamma & 0  & 0 & s_\gamma\\
 0  & 0 & 0 & 1 & 0 & 0 \\ 0 & 0 & 0 & 0 & 1 & 0 \\  0 & 0 &- s_\gamma & 0 & 0
 & c_\gamma   \end{array} 
\right)$
\vspace{5mm}

  $ \mathcal{P}(\nu_\mu \rightarrow\nu_e)$ 
is obtained from the 6x6 U matrix and the neutrino mass differences
$\delta m_{ij}^2=m_i^2-m_j^2$ for a neutrino beam with energy $E$ and baseline
$L$ by
\beq
\label{Pue5}
 \mathcal{P}(\nu_\mu \rightarrow\nu_e) &=& Re[\sum_{i=1}^{6}\sum_{j=1}^{6}
U_{1i}U^*_{1j}U^*_{2i}U_{2j} e^{-i(\delta m_{ij}^2/E)L}] \; ,
\eeq
an extension of the 4x4\cite{lsk14,lsk15} theory with one serile neutrino, 
which used the  3x3 formalism of Ref\cite{as97}, to a 6x6 matrix 
formalism\cite{tg07}.
From Eq(\ref{Uform3}), multiplying the 12 6x6 $O$ matrices, we obtain the 
matrix U. With $\delta_{CP}$=0, $U^*_{ij}=U_{ij}$, so we only need $U_{1j},U_{2j}$,
with the numbers in Eqs(\ref{U1j2},\ref{U2j}) given in Ref\cite{lskz15}.

\beq
\label{U1jz}
  U_{11}&=&.821 ca{\rm \;}cb{\rm \;}cg  \nonumber \\
  U_{12} &=& cg ((.554 ca - .821 sa^2) cb - .821 ca{\rm \;}sb^2) 
- .821 ca{\rm \;} cb{\rm \;}sg^2 \nonumber \\
 U_{13}&=&cg ((.15 ca-.554 sa^2-.821ca{\rm \;}sa^2)cb-(.554 ca - .821 sa^2)sb^2
\nonumber \\
      && -.821ca{\rm \;}cb{\rm \;}sb^2) - .821 ca{\rm \;}cb{\rm \;}cg{\rm \;}
sg^2 - ((.554 ca - .821 sa^2) cb -.821 ca{\rm \;}sb^2)sg^2 \nonumber \\
 U_{14} &=&cg(cb(.15sa +.554 ca{\rm \;}sa + .821 ca^2{\rm \;}sa)-.821ca{\rm \;}
cb^2{\rm \;}sb^2
\nonumber \\
   && -(.554 ca-.821 sa^2)cb{\rm \;}sb^2-(.15 ca-.554 sa^2-.821 ca sa^2)sb^2)
 - .821ca{\rm \;}cb{\rm \;}sg^2 cg^2\nonumber \\
  &&-cg((.554ca-.821sa^2) cb -.821ca {\rm \;}sb^2)sg^2
-(cb (.15 ca - .554 sa^2 - .821 ca sa^2)\nonumber \\
  && - .821ca{\rm \;}cb{\rm \;}sb^2 -(.554 ca-.821 sa^2) sb^2)sg^2 
\nonumber 
\eeq
\newpage

\beq
\label{U1j2}
U_{15} &=&cg(.821ca{\rm \;}sb{\rm \;}cb^3+(.15sa+.554ca{\rm \;}sa+
.821 ca^2{\rm \;}sa)sb \nonumber \\
   && +(.554 ca-.821 sa^2)cb^2{\rm \;}sb+(.15ca-.554sa^2-.821ca{\rm \;}sa^2)
cb{\rm \;}sb)
\nonumber \\
 && -.821ca{\rm \;}cb{\rm \;}cg^3 sg^2 -cg^2(cb(.554ca-.821 sa^2)-.821 sb^2)sg^2
\nonumber \\
  &&-cg(cb (.15 ca-.554 sa^2-.821ca{\rm \;}sa^2)-.821 ca{\rm \;}cb{\rm \;}sb^2
 \nonumber \\ 
   && -(.554 ca-.821 sa^2)sb^2 sg^2 -(cb(.15sa+.554ca{\rm \;} sa +
.821ca^2 sa) - .821 ca{\rm \;}cb^2{\rm \;}sb^2  \nonumber \\
   && -cb(.554ca -.821sa^2)sb^2+(.15ca-.554sa^2-.821ca{\rm \;}sa^2)sb^2) sg^2
 \nonumber \\ 
 U_{16}&=& .821ca{\rm \;}cb{\rm \;}sg{\rm \;}cg^4+(.821ca{\rm \;}cb^3{\rm \;}sb
 + (.15sa+.554ca{\rm \;}sa+.821ca^2{\rm \;}sa)sb\nonumber \\
   && +cb^2 (.554 ca-.821 sa^2)sb + 
    cb(.15 ca -.554 sa^2-.821 ca{\rm \;} sa^2) sb) sg \nonumber \\
   &&+cg^3 ((.554 ca - .821 sa^2) cb - .821 ca{\rm \;}sb^2)sg +\nonumber \\
   && cg^2 (cb (.15 ca -.554 sa^2 - .821ca{\rm \;}sa^2)-.821ca{\rm \;}cb
{\rm \;}sb^2\nonumber \\
   && - (.554 ca - .821 sa^2) sb^2) sg\nonumber \\
   &&+cg(cb(.15 sa+.554ca{\rm \;}sa+.821 ca^2{\rm \;}sa)-.821ca{\rm \;}cb^2 sb^2
\nonumber \\
  && -cb(.554ca-.821sa^2)sb^2-(.15 ca-.554 sa^2-.821ca{\rm \;}sa^2)sb^2) sg
\; ,
\eeq

\beq
\label{U2j}
  U_{21}&=& -.484ca{\rm \;}cb{\rm \;}cg \nonumber \\
  U_{22}&=&cg(.527ca+.484 sa^2)cb-.821ca{\rm \;}sb^2)+
.484ca{\rm \;}cb{\rm \;}sg^2
\nonumber \\
  U_{23}&=& cg((.699ca-.527sa^2+.484ca{\rm \;}sa^2)cb-(.527ca+.484sa^2)sb^2 
+.484ca{\rm \;}cb{\rm \;}sb^2)\nonumber \\
  && +.484ca{\rm \;}cb{\rm \;}cg{\rm \;}sg^2-((.527ca + .484sa^2)cb
+.484 ca{\rm \;}sb^2)*sg^2 
\nonumber \\
  U_{24}&=& cg(cb(.699 sa+.527ca{\rm \;}sa-.484ca^2{\rm \;}sa)+
.484ca{\rm \;}cb^2{\rm \;}sb^2 \nonumber \\
   && -(.527ca +.484sa^2)cb{\rm \;}sb^2 -(.699ca-.527sa^2+.484ca{\rm \;}sa^2) 
sb^2) +.484ca{\rm \;}cb{\rm \;}sg^2{\rm \;}cg^2 \nonumber \\
   &&-cg((.527 ca +.484sa^2)cb+.484ca{\rm \;}sb^2)sg^2-(cb(.69 ca-.527sa^2+
 .484ca{\rm \;}sa^2) +\nonumber \\ 
   && .484ca{\rm \;}cb{\rm \;}sb^2-(.527ca + .484 sa^2)sb^2)sg^2 
 \nonumber \\
  U_{25}&=& cg(-.484 ca{\rm \;}sb{\rm \;}cb^3 +(.699sa+.527ca{\rm \;}sa
-.484ca^2{\rm \;}sa)sb\nonumber \\
   && +(.527ca+.484sa^2)cb^2{\rm \;}sb +(.699ca -.527sa^2+.484 ca{\rm \;}sa^2) 
cb{\rm \;}sb)+.484ca{\rm \;}cb{\rm \;}cg^3{\rm \;}sg^2 \nonumber \\
   &&-cg^2(cb(.527ca+.484sa^2)+.484ca{\rm \;}sb^2)sg^2
-cg(cb(.699ca-.527sa^2+.484ca{\rm \;}sa^2)+
\nonumber \\
    && .484ca{\rm \;}cb{\rm \;}sb^2-(.527ca+.484sa^2)sb^2)sg^2
-(cb(.699sa+.527ca{\rm \;}sa-.484ca^2{\rm \;}sa)+ \nonumber \\
   &&.484ca{\rm \;}cb^2{\rm \;}sb^2 -cb(.527ca +.484sa^2)sb^2
+(.699ca-.527sa^2+.484 ca{\rm \;}sa^2)sb^2)sg^2 \nonumber \\
 U_{26}&=& -.484ca{\rm \;}cb{\rm \;}sg{\rm \;}cg^4+
(-.484ca{\rm \;}cb^3{\rm \;}sb + (.699 sa + .527 ca sa-.484ca^2{\rm \;}sa)sb
\nonumber \\
   && +cb^2 (.527ca + .484sa^2)sb+ cb(.699ca-.527sa^2+.484ca{\rm \;}sa^2)sb)sg
\nonumber\\
   && +cg^3((.527ca+.484 sa^2)cb + .484 ca{\rm \;}sb^2)sg +\nonumber\\
   && cg^2(cb(.699ca-.527sa^2+.484ca{\rm \;}sa^2)+.484 ca{\rm \;}cb{\rm \;}sb^2
    - (.527 ca + .484 sa^2) sb^2) sg \nonumber\\
   &&+cg(cb(.699sa+.527ca{\rm \;}sa-.484ca^2 sa)+.484ca{\rm \;}cb^2{\rm \;}sb^2
   - cb(.527ca +.484sa^2)sb^2 \nonumber\\
   &&-(.699 ca-.527sa^2 +.484 ca{\rm \;}sa^2)sb^2)sg \; .
\eeq

\newpage

\section{$\mathcal{P}(\nu_\mu \rightarrow \nu_e)$ For equal sterile
 neutrino masses}
Assuming that all three sterile neutrinos have the same mass, sterile-active
neutrino mass differences are $\delta m_{4j}^2=m_4^2-m_j^2 \simeq .9 (eV)^2$,
with $\delta m_{4j}^2$ taken from the best fit to neutrino oscillation 
data\cite{mini13} (see Ref\cite{mini13} for references to earlier experiments),
from Eq(\ref{Pue5}) $\mathcal{P}(\nu_\mu \rightarrow \nu_e)$ is 
\beq
\label{Pue6}
\mathcal{P}(\nu_\mu \rightarrow \nu_{e}) &=&Re[U_{11}U_{21}[ U_{11}U_{21}+
 U_{12}U_{22} e^{-i\delta L}+ U_{13}U_{23} e^{-i\Delta L}+ \nonumber \\
  && (U_{14}U_{24}+U_{15}U_{25} +U_{16}U_{26}) e^{-i\gamma L}]+ \nonumber \\
  &&  U_{12}U_{22}[ U_{11}U_{21}e^{-i\delta L}+ U_{12}U_{22} + U_{13}U_{23} 
e^{-i\Delta L}+ \nonumber\\
  && (U_{14}U_{24}+U_{15}U_{25} +U_{16}U_{26}) e^{-i\gamma L}]+ \nonumber \\
  &&  U_{13}U_{23}[ U_{11}U_{21}e^{-i\Delta L}+ U_{12}U_{22}e^{-i\Delta L}
 \nonumber \\
  && + U_{13}U_{23} 
    +(U_{14}U_{24}+U_{15}U_{25} +U_{16}U_{26}) e^{-i\gamma L}]+ \nonumber \\
   &&  (U_{14}U_{24}+U_{15}U_{25}+U_{16}U_{26})[(U_{11}U_{21}+ U_{12}U_{22}
  \nonumber \\
   &&+ U_{13}U_{23})e^{-i\gamma L}+U_{14}U_{24}+U_{15}U_{25}+U_{16}U_{26}]] \; ,
\eeq 
with $\delta=\delta m_{12}^2/2E,\; \Delta=\delta m_{13}^2/2E,\; \gamma=
\delta m_{jk}^2/2E$ (j=1,2,3;k=4,5,6). The neutrino mass differences are 
$\delta m_{12}^2=7.6 \times 10^{-5}(eV)^2$, $\delta m_{13}^2 = 2.4\times 10^{-3} 
(eV)^2$; and  $\delta m_{jk}^2 (j=1,2,3;k=4,5,6) =0.9 (eV)^2$\cite{mini13}.
\vspace{3mm}

 From Eq(\ref{Pue6})
\beq
\label{Pue7}
 \mathcal{P}(\nu_\mu \rightarrow\nu_e) &=& U_{11}^2 U_{21}^2+
 U_{12}^2 U_{22}^2+ U_{13}^2 U_{23}^2+  \nonumber \\
  && (U_{14}U_{24}+U_{15}U_{25}+U_{16}U_{26})^2 + \nonumber \\
  &&  2U_{11} U_{21} U_{12} U_{22} cos\delta L + \\
  && 2(U_{11} U_{21} U_{13} U_{23}+ U_{12} U_{22} U_{13} U_{23})cos\Delta L+
\nonumber \\
  &&2(U_{14}U_{24}+U_{15}U_{25}+U_{16}U_{26}) \nonumber \\
  &&(U_{11} U_{21}+U_{12} U_{22}+U_{13} U_{23})cos\gamma L \nonumber \; .
\eeq 
\vspace{3mm}

  Note that $\alpha\simeq 9.2^o$ from a recent analysis of MiniBooNE data, 
which was used in a recent study of $\mathcal{P}(\nu_\mu \rightarrow\nu_e)$ 
with one sterile neutrino\cite{lsk14,lsk15}. The figure below shows 
$\mathcal{P}(\nu_\mu \rightarrow \nu_e)$ with $\alpha=\beta=\gamma= 0^o$,
giving the results of a recent 3x3 S-mtrix calculation\cite{lsk15}.
In Fig. 6, for the other curves, the sterile-active mixing angle 
$\alpha=9.2^o$, while $\beta$ and  $\gamma$ are chosen to be $9.2^o$ and 
$20^o$ to compare the 6x6 to the previous 3x3 results. 
\newpage

\newpage
  Using Eq(\ref{Pue7}), one finds $\mathcal{P}(\nu_\mu \rightarrow \nu_e)$ for 
the 6x6 vs 3x3 theories:

\vspace{5cm}

\begin{figure}[ht]
\begin{center}
\epsfig{file=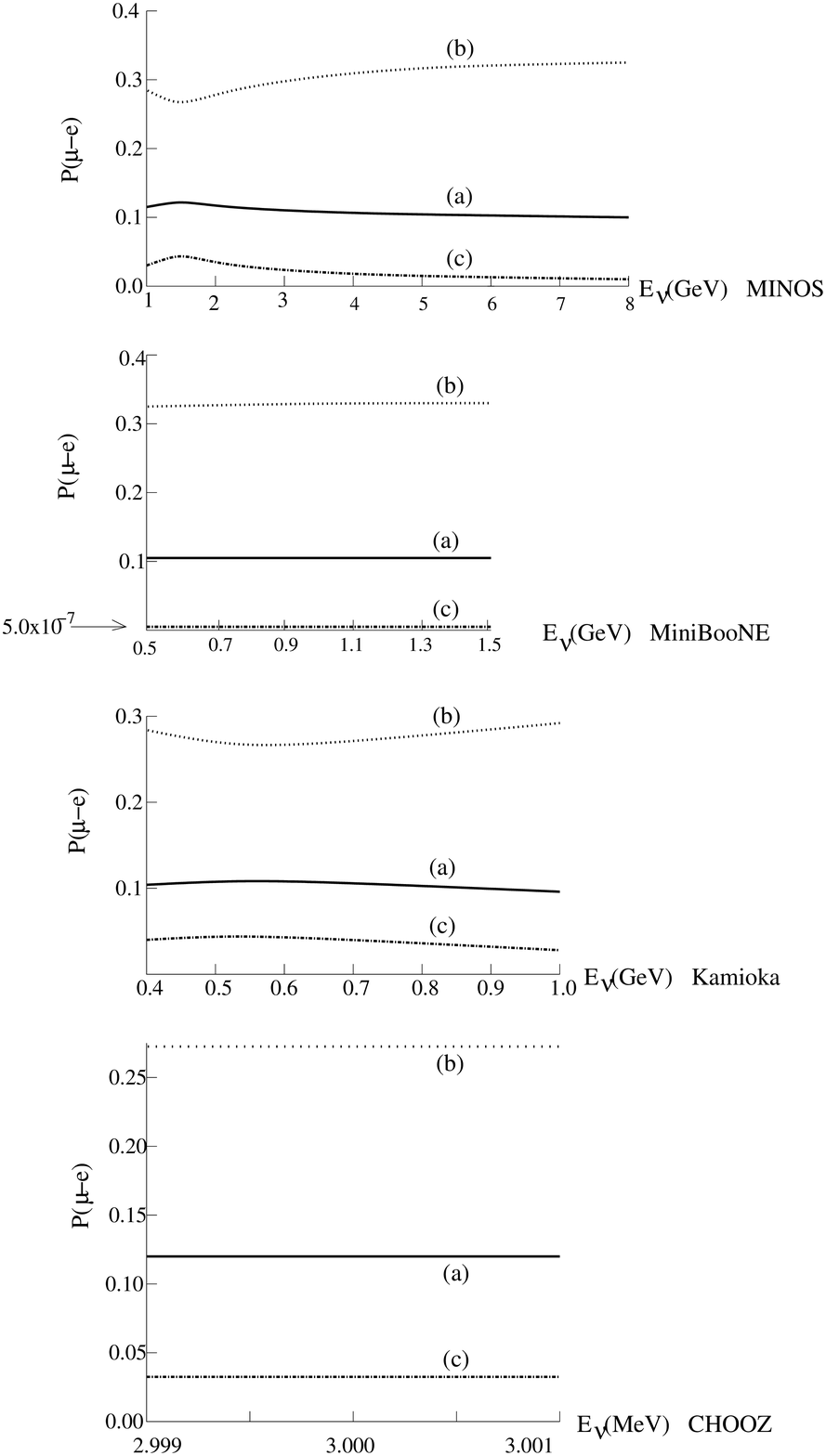,height=12cm,width=10cm}
\end{center}
\caption{$\mathcal{P}(\nu_\mu \rightarrow\nu_e)$ 
for MINOS(L=735 km), MiniBooNE(L=500m), JHF-Kamioka(L=295 km), and 
CHOOZ(L=1.03 km). (a) solid, for $\alpha=\beta=\gamma$=$9.2^o$;  
(b) dashed, for $\alpha, \beta, \gamma$ =$9.2^o$, $20^o$, $20^o$; 
(c) dash-dotted curve for $\alpha=\beta=\gamma$=$0^o$ giving the 3x3 result .}
\end{figure}
\vspace{3mm}
\newpage
\section{Conclusions}

In this review with sterile and active neutrinos the $\nu_\mu \rightarrow\nu_e$
transition probability $\mathcal{P}(\nu_\mu \rightarrow\nu_e)$ for only active
and one, two, three sterile neutrinos, with a variety of parameters associated 
with the transition probability, were discussed. There is now evidence for the
existence of two sterile neutrinos, but this is still uncertain. One
motivation for this review is to help extract parameters from future neutrino
oscillation experiments. 

At the present time there is no experimental evidence for three sterile
neutrinos. From studies of the Cosmic Microwave Background Radiation,
e.g. WMAP\cite{WMAP}, one knows that about 23 percent of matter in the
universe is Dark Matter, which consists of particles that have no interaction
except gravity. Since sterile neutrinos also only have a gravitational 
interaction, if the particles of Dark Matter are Fermions (quantum spin 1/2)
they might be massive sterile neutrinos. Thus a third sterile neutrino would
exist.

\Large
{\bf Acknowledgements}
\vspace{3mm}

\normalsize
This work was carried out while LSK was a visitor at Los Alamos
National Laboratory, Group P25. The author thanks Dr. William Louis for 
information about recent and future neutrino oscillation experiments,
and Dr. Terrance Goldman for advice on the mixing angles.

\end{document}